\documentclass[aps,preprint]{revtex4}%
\usepackage{amsfonts}
\usepackage{amsmath}
\usepackage{amssymb}
\usepackage{graphicx}%
\begin{document}
\title{ Jamming transition in emulsions and granular materials}
\author{H. P. Zhang}
\affiliation{Physics Department, City College of New York}
\author{H. A. Makse}
\affiliation{Physics Department and Levich Institute, City College of New York}

\begin{abstract}
We investigate the jamming transition in packings of emulsions and granular
materials via molecular dynamics simulations. The emulsion model is composed
of frictionless droplets interacting via nonlinear normal forces obtained
using experimental data acquired by confocal microscopy of compressed
emulsions systems. Granular materials are modeled by Hertz-Mindlin deformable
spherical grains with Coulomb friction. In both cases, we find power-law
scaling for the vanishing of pressure and excess number of contacts as the
system approaches the jamming transition from high volume fractions. We find
that the construction history parametrized by the compression rate during the
preparation protocol has a strong effect on the micromechanical properties of
granular materials but not on emulsions. This leads the granular system to jam
at different volume fractions depending on the histories. Isostaticity is
found in the packings close to the jamming transition in emulsions and in
granular materials at slow compression rates and infinite friction.
Heterogeneity of interparticle forces increases as the packings approach the
jamming transition which is demonstrated by the exponential tail in force
distributions and the small values of the participation number measuring
spatial localization of the forces. However, no signatures of the jamming
transition are observed in structural properties, like the radial distribution
functions and the distributions of contacts.

\end{abstract}

\pacs{81.05.Rm}
\maketitle

\section{Introduction}

A variety of systems such as granular materials, compressed emulsions, or
molecular glasses, exhibit a non-equilibrium transition from a fluid-like to a
solid-like state. When the constituent particles of such systems are so
crowded that they are in close contact with one another, the whole system
experiences a sudden dynamical arrest, which is referred to as the glass or
jamming transition \cite{Liu1,hernan2003,coniglio}. A jammed system is a
many-body system blocked in a configuration far from equilibrium, from which
it takes too long a time to relax in the laboratory time scale. Jammed systems
are usually random packings of particles, which have been the starting point
for many studies of liquids and glasses
\cite{Bernal,scott,onoda,Torquato,coniglio}. A theoretical approached to treat
these systems has been developed by Edwards and coworkers
\cite{edwards1994,mehta} who have proposed a statistical mechanics description
of such jammed system, in which thermodynamic quantities are computed as flat
average over jammed configurations \cite{kurchan,bkls,hernan-nat}.

In a recent paper \cite{hernan1} we found that the jamming transition in
packings of granular materials and emulsions can be viewed as a phase
transition at the critical concentration of random close packing (RCP) for
frictionless emulsions or random loose packings (RLP) for frictional grains.
The transition is characterized by a power behavior and the corresponding set
of exponents describing the stress, coordination number and elastic moduli as
the system approaches the jamming transition. This idea has also been
investigated for packings of frictionless particles \cite{ohern1,ohern2} and
colloids \cite{trappe}\textit{. } Here we elaborate further on the results of
\cite{hernan1} and present our detailed computer simulation studies about many
aspects of the jamming transitions in these systems.

We consider packings of spherical deformable particles interacting via
friction and frictionless forces. Granular packings are characterized by
elastic Hertz-Mindlin forces \cite{Landau,mindlin,Hernan2004} with Coulomb
friction. Concentrated emulsion systems are modeled as a collection of
frictionless deformable droplets interacting via a nonlinear normal force law
which is obtained from the analysis of experimental data acquired using
confocal microscopy in \cite{Jasna}.

We first introduce a protocol to generate a series of packings near the
jamming transition and beyond for both emulsions and granular materials. Then
we study how the construction history of the packings affects the jamming
transition. The existence of frictional tangential forces is one of many
generic properties of granular materials. In the elastic Hertz-Mindlin model
the tangential force is proportional to the tangential displacement, and it is
truncated by a Coulomb cut-off. Both the elastic tangential force and Coulomb
friction are path-dependent. The path-dependence in the microscopic
interactions leads to the path-dependence of the granular packings at the
macroscopic level, an effect that has been shown in several classical
experiments \cite{scott,onoda,nowak}. In our simulations we find that the
construction history affects the properties at the jamming transition in the
granular packings through their direct impact on the tangential forces, while
in the case of emulsions we do not find any path-dependence.

Next, we fully characterize the evolution of the micromechanical and
microstructural properties, such as the stress, the coordination number and
isostaticity, the radial distribution function, the probability distributions
of normal and tangential forces and interparticle contacts, the existence of
force chains, the participation number and plasticity index, as the system
approaches the jamming transition. We found that the heterogeneity of
interparticle forces increases as the packings approach the jamming transition
which is demonstrated by the exponential tail in force distributions and the
small values of the participation number. However, no clear signatures of the
jamming transition are observed in structural properties, like the radial
distribution functions and the distributions of contacts.

The paper is organized as follows: in Section \textbf{II} we discuss the
microscopic force models used in our simulations of granular materials and
emulsions. In Section \textbf{III} we give the details of the numerical
protocol to generate packings near the jamming transition and beyond and the
various measurements to be performed. The main results of this work are given
in Section \textbf{IV} and we summarize our results in Section \textbf{V}.

\section{Microscopic model}

In this section we discuss the microscopic models of inter-particle forces for
granular materials and emulsions. We first briefly review the standard
Hertz-Mindlin model for granular materials; more details can be found in
\cite{ctctmerch,Landau,Hernan2004}. Then we describe the linear "Princen
model" \cite{Princen} for the inter-particle forces in emulsions and its
nonlinear modification according to experimental data on deformation in
concentrated emulsion systems \cite{Jasna}.

\subsection{Contact mechanics in granular materials}

We describe the microscopic interaction between grains by the nonlinear
Hertz-Mindlin normal and tangential forces. The normal force between two
contacting grains at position $\overrightarrow{x}_{1}$ and $\overrightarrow
{x}_{2}$ with uncompressed radii $R_{1}$ and $R_{2}$ is
\cite{ctctmerch,Landau}
\begin{equation}
F_{n}=\frac{2}{3}k_{n}R^{1/2}\xi^{3/2}, \label{gran_normal}%
\end{equation}
where $R$ is the geometric mean of $R_{1}$ and $R_{2}:$ $R=2R_{1}R_{2}%
/(R_{1}+R_{2})$, and $\xi$ is the normal overlap: $\xi=\left(  1/2\right)
\left[  R_{1}+R_{2}-\left\vert \overrightarrow{x}_{1}-\overrightarrow{x}_{2}
\right\vert \right]  .$ The normal force acts only in compression, i.e.
$F_{n}=0$ when $\xi<0.$ The effective stiffness $k_{n} =4G/(1-\nu)$ is defined
in terms of the shear modulus of the grains $G$ and the Poisson ratio $\nu$ of
the material the grains are made. From Eq. (\ref{gran_normal}), we can see
that the normal forces are completely determined by the geometrical
configuration of the packing and have nothing to do with the construction history.

The tangential contact force was first calculated by Mindlin \cite{mindlin}
for grains under oblique loading. In his model, the increment in tangential
force is
\begin{equation}
\Delta F_{t}=k_{t}(R\xi)^{1/2}\Delta s \label{gran-tan}%
\end{equation}
where $k_{t}=8G/(2-\nu),$ and the variable $s$ is defined such that the
relative tangential displacement between the two grain centers is $2s$.
Therefore the tangential force is obtained by integrating Eq. (\ref{gran-tan})
over the path taken by the particles in contact with the initial condition,
$F_{n}=0,F_{t}=0$ at $\xi=0,s=0,$ yielding:%

\begin{equation}
F_{t}=%
{\displaystyle\int\limits_{path}}
k_{t}(R\xi)^{1/2}ds\text{ }. \label{grain_tang_int}%
\end{equation}
As the tangential displacement increases, the elastic tangential force $F_{t}$
reaches its limiting value given by a Coulomb cut-off for granular materials,
which is
\begin{equation}
F_{t}=\mu F_{n}. \label{cuttoff}%
\end{equation}
The Coulomb cut-off adds a second source of path-dependence to the problem.

From the discussion above, one can see that a straightforward way to affect
the tangential force is to change the frictional coefficient $\mu$, which
controls the maximum value of the ratio between tangential and normal force.
But the elastic tangential force, Eq. (\ref{gran-tan}), is even more
intimately related to the tangential displacement, which is mainly determined
by the construction history.

\subsection{Force law for emulsion}

Emulsions are a class of material which is both industrially important and
exhibits very interesting physics \cite{Princen}. They belong to a wider
material class of colloids in that they consist of two immiscible phases one
of which is dispersed into the other, the continuous phase. Both of the phases
are liquids and their interface is stabilised by the presence of
surface-active species. Emulsions are composed of droplets of a liquid (for
instance silicone oil) stabilized by a surfactant (like sodium dodecylsulphate
) in a continuum phase (such as a water and glycerol mixture) \cite{Jasna}.
Being composed of only liquids, emulsion droplets interact with each other
only via normal forces with no solid friction between them. The determination
of an accurate force model for the compression of two droplets is by no means
trivial, but it can be simplified in certain limits. For small deformation
with respect to the droplets surface area the Laplace pressure remains
unchanged and all energy of the applied stress is presumed to be restored in
the deformation of the surface. Then the normal repulsive force between two
spherical droplets in contacts with uncompressed radii $R_{1}$ and $R_{2}$ can
be calculated at the microscopic level as:%

\begin{equation}
f=\frac{\sigma}{R}A_{mea} \label{princen}%
\end{equation}
This is known as the Princen model \cite{Princen}, where $\sigma$ is the
interfacial tension of droplets, $f$ is the normal force between two droplets,
$R$ is the geometric mean of $R_{1}$ and $R_{2},$ and $A_{mea}$ is the
measured contact area between the droplets. As in the granular materials, the
normal force acts only in compression.

In \cite{Jasna} a linear model was used to relate the area of deformation with
the overlap between the spheres $\xi$, resulting in a linear spring model for
the force-law between the droplets. More detailed calculations \cite{morse}
and numerical simulations \cite{lacasse} show that the interdroplet forces in
emulsions are better represented by a nonlinear spring $f\propto\xi^{\alpha}$
with the exponent $\alpha$ between $1$ and $1.5$ and, more importantly,
depending on the number of contact forces acting on the droplet.

\begin{figure}
[ptb]
\begin{center}
\includegraphics[
height=2.4811in,
width=6.8061in
]%
{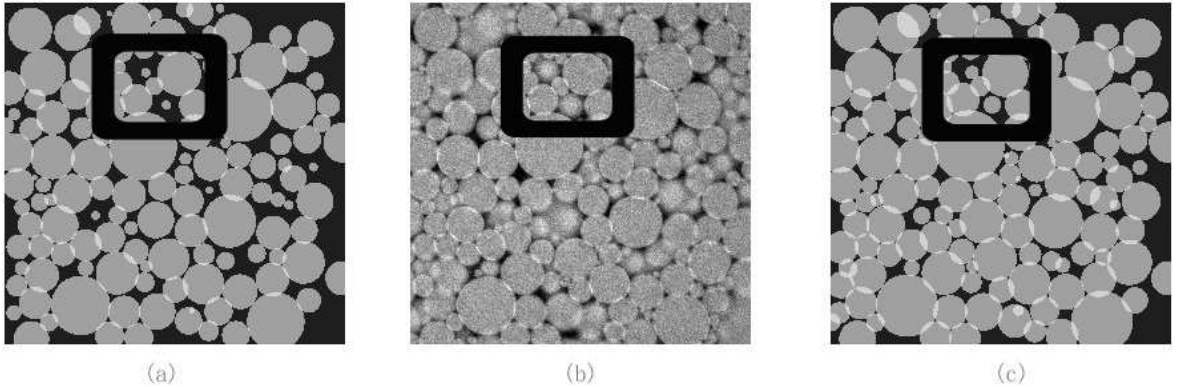}%
\caption{Image slice of emulsion packing from confocal microscope (b);
reconstructed image before (a) and after Monte Carlo annealing (c).}%
\label{emulpic}%
\end{center}
\end{figure}

Recently, Bruji\'{c} \textit{et al.}\textbf{ }\cite{Jasna} used confocal
microscopy to study a compressed polydisperse emulsion. This system consists
of a dense packing of emulsion oil droplets, with a sufficiently elastic
surfactant stabilising layer to mimic solid particle behaviour, suspended in a
continuous phase fluid. By refractive index matching of the two phases they
obtained a 3D image of the packing structure by using confocal microscopy for
the imaging of the droplet packings at varying external pressures, i.e. volume
fractions. The key feature of this optical microscopy technique is that only
light from the focal plane is detected. Thus 3D images of translucent samples
can be acquired by moving the sample through the focal plane of the objective
and acquiring a sequence of 2D images. One 2D image obtained experimentally is
shown in Fig. \ref{emulpic}a.

The emulsion system, stable to coalescence and Ostwald ripening, consisted of
Silicone oil droplets ($\eta=10$cS) in a refractive index matching solution of
water ($w_{t}=51\%$) and glycerol ($w_{t}=49\%$), stabilised by 20$mM$ sodium
dodecylsulphate (SDS) upon emulsification and later diluted to below the
critical micellar concentration (CMC$=13mM$) to ensure a repulsive
interdroplet potential. The droplet phase is fluorescently dyed using Nile
Red, prior to emulsification. The control of the particle size distribution,
prior to imaging, is achieved by applying very high shear rates to the sample,
inducing droplet break-up down to a radius mean size of $3.4\mu m$. This
system is a modification of the emulsion reported by Mason \textit{et al.}
\cite{mason} to produce a transparent sample suitable for confocal microscopy.

\begin{figure}
[ptb]
\begin{center}
\includegraphics[
height=3.4143in,
width=4.0603in
]%
{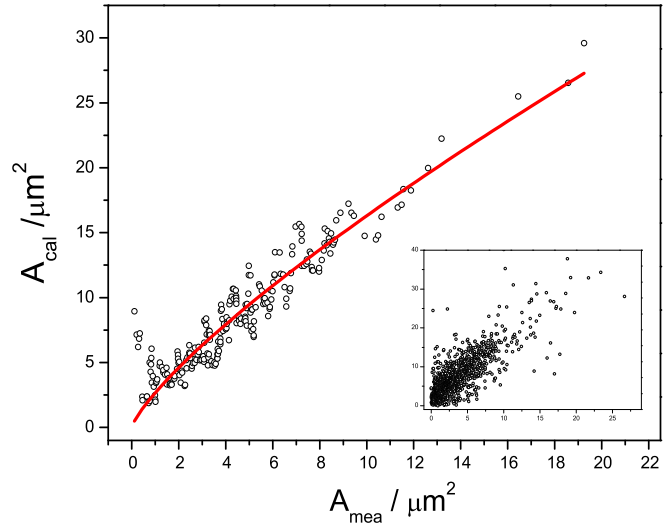}%
\caption{$A_{cal}$ is plotted against $A_{mea}$ after Monte Carlo annealing;
the red line is a fit of . In the inset, $A_{cal}$ is plotted against
$A_{mea}$ for all forces before Monte Carlo annealing.}%
\label{force_law_emul}%
\end{center}
\end{figure}

In these images, the area of contacts, the droplets and the aqueous background
differ by the darkness in an 8-bit gray-scale image such as the one displayed
in Fig. \ref{emulpic}a which has an average darkness 210, 90, and 30,
respectively. The result of the image analysis carried out by Bruji\'{c}
\textit{et al.} is a set of contact areas, $A_{mea},$ along with the
undeformed radii $(R_{1},R_{2})$ and positions $(\vec{x}_{1},\vec{x}_{2})$ of
two droplets giving rise to each contact.
Using the obtained information of $R_{i}$ and $\overrightarrow{x}_{i}$ and
ignoring the droplet deformations$,$ we can reconstruct the images as shown in
Fig. \ref{emulpic}b and calculate a geometric overlapping area
\begin{equation}
A_{cal}=\pi R\xi, \label{EQ.2}%
\end{equation}
where $\xi$ is the overlapping. For all the 1439 contacts obtained in the
experiments, $A_{cal}$ is plotted against $A_{mea}$ in the inset of Fig.
\ref{force_law_emul}, from which we can see that $A_{cal}$ is generally
different from $A_{mea}$, indicating that there exist deviations from the
linear force law Eq. (\ref{princen}). This fact provides a direct measure of
the effects of anharmonicity of interaction between the droplets surfaces.%

In order to extract the nonlinear dependence of the force law between the
force and the deformation characterized by $\xi$, we plot the calculated area
$A_{cal}$ from the reconstructed image versus the real area of deformation
$A_{mea}$. In order to achieve this, we need to improve the estimation of the
droplet centers and radii by minimizing the difference between the
reconstructed image and the original one. In this minimization, $R_{i}$ and
$\vec{x}_{i}$ are the changing variables and we use a Monte Carlo annealing
method to find the optimum set of parameters. The corresponding image after
the minimization is shown in Fig. \ref{emulpic}c, which is closer to the
original image than the reconstructed image shown in Fig. \ref{emulpic}b as
indicated by rectangular region in the figure. Finally, we choose only those
contacts for which $R_{i}$ and $\vec{x}_{i}$ change less than 5\% before and
after the minimization, and plot $A_{cal}$ versus $A_{mea}$ in the main panel
of Fig. \ref{force_law_emul}. The plot shows a clear trend which can be fitted
as
\begin{equation}
A_{cal}=2.67\left(  A_{mea}\right)  ^{0.78}. \label{eq3}%
\end{equation}
Combining Eqs. (\ref{princen}), (\ref{EQ.2}) and (\ref{eq3}), we obtain the
nonlinear force-law for emulsion droplets:
\begin{equation}
f=1.23~\sigma~\widetilde{R}^{0.28}~\xi^{1.28}. \label{emulforce}%
\end{equation}
In our simulations, Eq. (\ref{emulforce}), which takes in account partially
the effects of the anharmonicity of interaction between the droplets, is used.
We also use a linear spring force law $f=4\pi\sigma\xi$, which completely
neglects the anharmonicity of interaction. These two forces give similar
results in the quantities we measured. In this paper, we only present the
results using Eq. (\ref{emulforce}).%

\section{Simulations, protocol and measurements}

In this section, we first describe the method of molecular dynamics
simulations. Then we show the protocol to generate packings approaching the
jamming transition, and in the third part we define various quantities we use
to characterize the jamming transition.

\subsection{Molecular dynamics simulation}

We follow the Discrete Element Method (DEM) or Molecular Dynamics developed by
Cundall and Strack \cite{cundall} and solve the Newton equations for an
assemble of particles with zero-gravity, which interact via the microscopic
models discussed in Section II. DEM employs a time-stepping, finite-difference
approach to solve the Newton equations of motion simultaneously for every
particle in the system:%
\begin{equation}
\overrightarrow{F}=m\frac{d^{2}\overrightarrow{X}}{dt^{2}} \label{smlt_tran}%
\end{equation}

\begin{equation}
\overrightarrow{M}=I\frac{d^{2}\overrightarrow{\theta}}{dt^{2}}
\label{smlt_rot}%
\end{equation}
where $\overrightarrow{F}$ and $\overrightarrow{M}$ are the total force and
torque acting on a given particle, $m$ and $I$ are the mass and moment of
inertia, $\vec{X}$ and $\vec{\theta}$ are the position and angle of the
particles respectively. The numerical solution of Eq. (\ref{smlt_tran}) and
Eq. (\ref{smlt_rot}) are obtained by integration, assuming constant velocities
and accelerations for a given time step. We choose the time step to be a
fraction of the time that it takes for sound wave to propagate on one grain
(or droplet). A global damping proportional to the translational velocity was
used to dissipate energy in both cases. The damping simulates the drag force
of the droplets in the suspending medium. More details about simulations can
be found in \cite{Hernan2004}.

In the simulations of emulsions, the system consists of 10000 droplets, which
are all $2$ $\mu m$ in diameter. Droplets interact via the normal force given
by Eq. (\ref{emulforce}) with the interfacial tension $\sigma=9.8\times
10^{-3}$N/m and the density $\rho=10^{3}$ kg/m$^{3}$. The granular system is
composed of 10000 glass beads of equal size (radius=$0.1$mm) interacting via
the Hertz-Mindlin forces, Eqs. (\ref{gran_normal}) and (\ref{grain_tang_int})
and the Coulomb force Eq. (\ref{cuttoff}). The microscopic parameters defining
the glass beads are the shear modulus: $G=29$Gpa, the Poisson's ration:
$\nu=0.2,$ the friction coefficients $\mu=0.3$ and the density: $\rho
=2\times10^{3}$ kg/m$^{3}$.

\subsection{Numerical protocol}

\begin{figure}
[ptb]
\begin{center}
\includegraphics[
height=2.8911in,
width=3.7118in
]%
{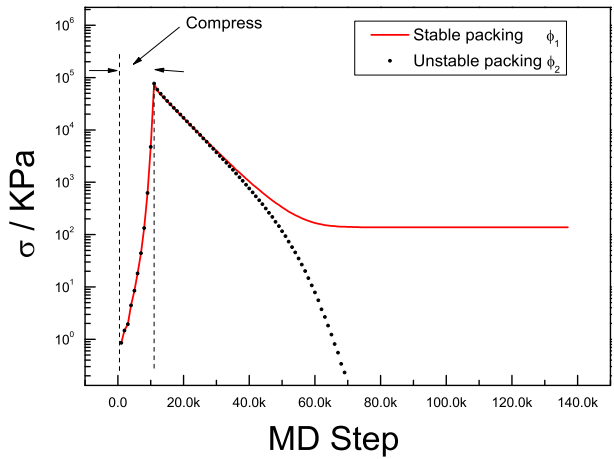}%
\caption{Time evolutions of pressure for two packings in simulation of
granular material. Line represents a packing with $\phi_{1}>\phi_{c}$ and
circles represent a packing with $\phi_{2}<\phi_{c}$, where $\phi_{1}=\phi
_{2}+1.6\times10^{-5}$.}%
\label{criticality}%
\end{center}
\end{figure}

We start our simulations from a set of non-overlapping particles located at
random positions in a periodically repeated cubic box with an initial volume
fraction around $\phi\approx0.2$. The box is compressed isotropically by
constant compression rate $\gamma$ until a given volume fraction, $\phi$, is
reached. Then the compression is turned off and the system is allowed to relax
with constant volume until it reaches a stable state, which means that the
pressure of system remains unchanged over a period of time (usually
$5\times10^{5}$ MD steps). This protocol can generate packings with different
volume fractions. From simulations, we find that there is a critical volume
fraction $\phi_{c},$ below which a jammed packing with nonzero pressure can
not be obtained. This fact is illustrated in Fig. \ref{criticality}, where the
time evolution of the pressure in this compress-and-relax process for two
packings are shown. The packing with $\phi_{1}>$ $\phi_{c}$ stabilizes at
none-zero pressures but the pressure decreases very fast to zero for $\phi
_{2}<$ $\phi_{c}$, even though $\phi_{1}$ and $\phi_{2}$ are different only by
$1.6\times10^{-5}$.%

Several other ways of preparing static packings of particles exist in the
literature. Conjugate gradient methods were used to study dense random packing
of frictionless particles by O'Hern \textit{et al. } \cite{ohern1}. A
pouring-ball method, which mimics the process of pouring balls into a
container under gravity, was used by Silbert \textit{et al. }
\cite{slibertgeo,slibertforce}. A servo mechanisms which adjusts the strain in
the system to equilibrate the packing at a given stress was used in
\cite{Hernan2004}. One of the advantages of the present protocol is that it
allows us to generate packings with different construction histories by using
different compression rates. Thus we can study the path-dependency systematically.

\subsection{Computation of characteristic quantities}

After generating the packings, we compute the following quantities to
characterize their micromechanical and structural properties:

\subsubsection{Pressure and coordination number}

The macroscopic stress tensor for point contacts in a volume $V$ is given by
\[
\sigma_{ij}=\frac{1}{V}\sum\limits_{k}R^{k}n_{i}^{k}F_{j}^{k}%
\]
where $V$ is the volume of the system, $\overrightarrow{R^{k}}$ is the vector
joining the center of two particles in contact $k$, $\overrightarrow{n_{k}%
}=\overrightarrow{R^{k}}/R^{k}$ and $\overrightarrow{F^{k}}$ is the total
force in contact $k$. The pressure $\sigma$ is the average of three diagonal
elements of the stress tensor, i.e. $\sigma=(\sigma_{11}+\sigma_{22}%
+\sigma_{33})/3.$ The coordination number, $Z$, is the average of contacts per
particle in contact network: $Z=\frac{2M}{N}$, where $M$ is the total contacts
and $N$ is the number of total particle in contact network. Due to the
non-gravity environment in our simulation, some floaters, which have zero
contacts and don't participate in the contact network, exist. We exclude them
when calculating the coordination number. We note the floaters were also
reported in Ref. \cite{ohern1}.

\subsubsection{Force distributions, force chains and participation number}

In both emulsion and granular packing, interparticle forces are highly
inhomogeneous. This is quantified by the probability distribution of the
normal and tangential forces calculated for emulsions and grains. Moreover,
photoelastic visualization experiments \cite{liu} and simulations
\cite{radjai,antony,hernan1} have shown that the contact forces are strongly
localized along "force chains" which carry most of the applied stress. To
quantify the degree of force localization, we consider the participation
number $\Gamma$ \cite{soares}:
\begin{equation}
\Gamma=\left(  M%
{\textstyle\sum\limits_{i=1}^{M}}
q_{i}^{2}\right)  ^{-1}. \label{gamma}%
\end{equation}
Here $M$ is the number of total contacts, and
\[
q_{i}=f_{i}/%
{\textstyle\sum\limits_{j=1}^{M}}
f_{j},
\]
where $f_{j}$ is the total force at contact $j.$ From the definition,
$\Gamma=1$ indicates $q_{i}=1/M,$ for all $q_{i}$ and a state with a spatially
homogeneous force distribution. On the other hand, in the limit of complete
localization, $\Gamma\approx1/M$, which is essentially zero in our simulation.

\subsubsection{Plasticity Index}

In order to quantitatively characterize the tangential forces in the granular
packings, we measure the plasticity index following \cite{slibertgeo}:
\begin{equation}
\Sigma=\frac{F_{t}}{\mu F_{n}}, \label{plasticity}%
\end{equation}
which has value between $0$ and $1$. The maximum means the contact is
"plastic" and the tangential force is the Coulomb friction; otherwise the
contact is elastic and the tangential force is the Mindlin elastic force.

\section{Results}

\subsection{Power-law scaling near jamming. Effect of construction history and
Isostaticity}

\begin{figure}
[ptb]
\begin{center}
\includegraphics[
height=3.8735in,
width=4.9139in
]%
{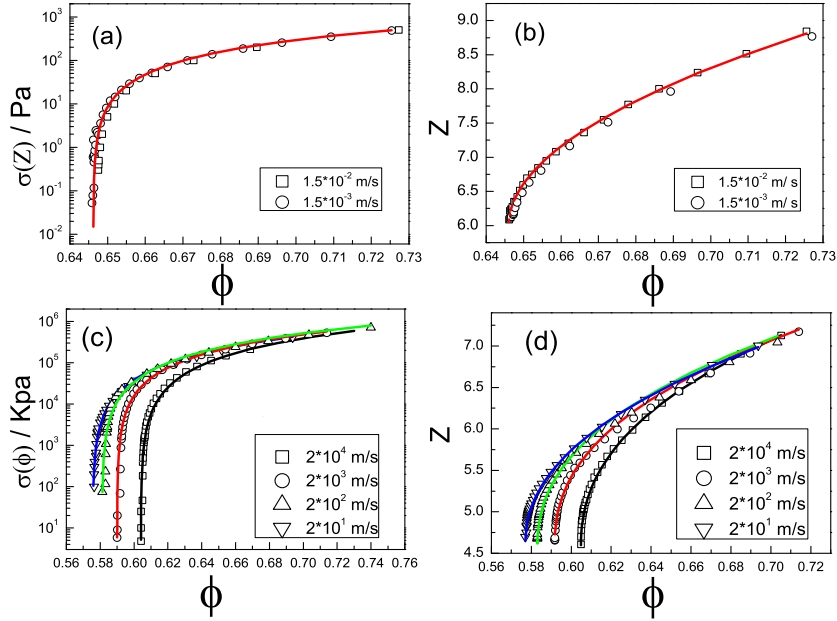}%
\caption{Power-law scalings in the packings of granular materials and
emulsions approaching the jamming transition. The symbols are the data from
the simulations and different symbol corresponds to different compression
rate. The lines are power-law fits. (a) $\sigma$ vs $\phi$ in emulsions, (b)
$Z$ vs $\phi$ in emulsions, (c) $\sigma$ vs $\phi$ in granular materials and
(d) $Z$ vs $\phi$ in granular materials.}%
\label{power-laws}%
\end{center}
\end{figure}

Using the protocol discussed above, we generate series of packings of
emulsions and granular materials approaching the jamming transition. In order
to study the effect of constructing history, different compression rates are
used to generate the packings. For the case of granular materials, we use four
compression rates: $2\times10^{4},2\times10^{3},2\times10^{2}$ and
$2\times10^{1}$m/s, while for emulsions we use two: $1.5\times10^{-2}$ and
$1.5\times10^{-3}$m/s. In Fig. \ref{power-laws} we plot the pressure, $\sigma
$, and coordination number, $Z$, of the resulting packings as a function of
volume fraction, $\phi$, where different symbols correspond to different
compression rates. Quantitative analysis of the data in Fig. \ref{power-laws}
shows that in both case the pressure $\sigma$ and extra contacts $Z-Z_{c}$
vanishes as power-laws as $\phi$ approaches the jamming transition at
$\phi_{c}$ as shown in \cite{hernan1}:
\begin{equation}
\sigma\sim(\phi-\phi_{c})^{\alpha} \label{power-1}%
\end{equation}
and
\begin{equation}
Z-Z_{c}\sim(\phi-\phi_{c})^{\beta}. \label{power-2}%
\end{equation}

The power-law fits are shown by lines in Fig. \ref{power-laws} and the
extracted fitting parameters including the critical volume fraction $\phi_{c}%
$, the critical (minimal) coordination number $Z_{c}$ and two power indexes
$\alpha$ and $\beta$ are listed in Table I.

As shown in Fig. \ref{power-laws}, the jamming transition for emulsions occurs
at $\phi_{c}=0.645$ and $Z_{c}=6$ regardless of the compression rate, i.e. the
construction history. We identify $\phi_{c}=0.645$ as random close packing
(RCP) volume fraction, which value has been reported many times before
\cite{hernan1,ohern1,slibertgeo}. The minimal coordination number $Z_{c}=6$
indicates that the packings near the jamming transition in emulsions are
isostatic \cite{Moukarzel,hernan1,ohern1,slibertgeo}.

A packing is isostatic \cite{Alex} when the number of contact forces equal the
number of force balance equations. In a 3D packing of perfectly smooth
(frictionless) particles there are $NZ/2$ unknown normal forces and $3N$ force
balance equations. This gives rise to a minimal coordination number needed for
mechanical stability as $Z_{c}=6$ in 3D (the isostatic limit). In the case of
packings of perfectly rough particles which is realized by frictional
particles with infinite friction $\mu\rightarrow\infty$--- notice that in this
case there are still tangential forces given by the Mindlin elastic component
Eq. (\ref{gran-tan})--- besides $NZ/2$ unknown normal forces and $3N$ force
balance equations, there are $NZ$ unknown tangential forces and $3N$ torque
balance equations. Thus the coordination number in the isostatic limit is
$Z_{c}=4$ for frictional packings in $3$D. In such isostatic packings, there
is possibly a unique solution for the forces between particles for a given
geometrical configuration, because the number of equations equals to the
number of unknowns. The existence of isostaticity is the foundation of recent
theories of stress propagation in granular materials
\cite{edward3,witten,ball}.

Thus, in our simulation the isostatic limit in the frictionless case is
approached as $\sigma\to0$ or $\phi\to\phi_{c}$. In other words, the
isostaticity appears in the limit of rigid balls, or when the rigidity of the
particles goes to infinity. This corresponds to the so-called marginal
rigidity state \cite{ball}.

Quite different from the results is emulsions, $\phi_{c}$ in frictional
granular materials with finite $\mu=0.3$ varies depending on the compression
rates from $0.576$ (slowest rate) to $0.604$ (fastest rate) as seen in Fig.
\ref{power-laws}. The critical coordination numbers in Table I for granular
materials are around $Z_{c}\approx4.5$. This value is above $Z_{c}=4$, the
value required by isostaticity. However, it does not mean that these
frictional granular packings with finite $\mu=0.3$ are not isostatic since the
prediction $Z_{c}=4$ is strictly valid for packings with $\mu\rightarrow
\infty$. We will elaborate more on this result below and discuss in more
detail this limit.%

\begin{table}[tbp] \centering
\begin{tabular}
[c]{|c|c|c|c|c|c|}\hline
System & Com. \ rate (m/s) & $\phi_{c}$ & $Z_{c}$ & $\alpha$ & $\beta$\\\hline
Emulsion, $\mu=0$ & $1.5\times10^{-2}\left(  10^{-3}\right)  $ & 0.645 &
6.01 & 1.25 & 0.51\\\hline
Granular, $\mu=0.3$ & $2\times10^{4}$ & 0.604 & 4.52 & 1.52 & 0.46\\\hline
Granular, $\mu=0.3$ & $2\times10^{3}$ & 0.590 & 4.53 & 1.46 & 0.45\\\hline
Granular, $\mu=0.3$ & $2\times10^{2}$ & 0.581 & 4.53 & 1.48 & 0.46\\\hline
Granular, $\mu=0.3$ & $2\times10^{1}$ & 0.576 & 4.54 & 1.52 & 0.47\\\hline
Granular, $\mu=\infty$ & $2\times10^{2}$ & 0.571 & 3.98 & 1.66 & 0.45\\\hline
\end{tabular}
\caption{Parameters in the power-law fittings near the
jamming transition Eqs. (\ref{power-1})  and (\ref{power-2}) for emulsions and
granular materials. }
\end{table}%

Next we try to understand the values of the power indexes. In the granular
systems, due to the Coulomb cut-off (with $\mu=0.3$), the intergranular normal
forces are always larger than the tangential forces. The main contribution to
the pressure is from the normal forces and not the tangential ones. Therefore,
for Hertzian balls we expect the following scaling:%

\[
\sigma\sim F_{n}\sim\xi^{3/2}\sim(\phi-\phi_{c})^{3/2},
\]
which implies that the exponent $\alpha$ in Eq. (\ref{power-1}) should be
equal to the 3/2 exponent of the normal force law in Eq. (\ref{gran_normal}).
The values of $\alpha$ obtained in our simulations for different compression
rates shown in Table I are all around $3/2$, supporting the above argument.
This argument should also hold in the case of emulsions since only normal
forces act between the droplets. In fact, we found $\alpha=1.25$ for emulsions
independent of the compression rate which is close to the exponent $1.28$ in
the force law Eq. (\ref{emulforce}). From Table I we find that the exponent
$\beta$ in Eq. (\ref{power-2}) are very close to $0.5$ for both emulsions and
grains, independent of the force law and also of the compression rate. This
numerical value and the independence of the force law were also reported by
O'Hern \textit{et al}. \cite{ohern2}. These authors provided a possible way to
understand $\beta=0.5$. In \cite{slibertgeo} it was shown that the radial
distribution function, $g(r)$, in packings near the jamming transition
displays a power-law behavior near $r=D$, where $D$ is the diameter of
spheres:%
\begin{equation}
g(r)\propto(1-r/D)^{-0.5}. \label{gr-power}%
\end{equation}
If one assumes an affine deformation upon compression, then one consequence of
Eq. (\ref{gr-power}) is that the coordination number should increase with the
power $\beta=0.5$. However, as shown below, we do not observe the power-law
region in $g(r)$ from our data. Therefore, to our opinion, the origin of
$\beta=0.5$ is still not clear. In \cite{Hernan2004} we provide more detailed
analysis of this argument assuming a narrow peak of the pair distribution
function at $r=D$. However, the predicted dependence of the coordination
number with pressure still disagree with the $\beta=0.5$ result.

Here we elaborate further on the topic and present an analogous derivation of
the exponent $\beta$. Let us assume that, in the limit of zero pressure, there
is a probability distribution $P(h)$ of gap sizes, $h$, between each ball and
its neighbors:%

\begin{equation}
P(h) = Z_{c} \delta(h) + d_{1} h^{-\xi}, \label{ph}%
\end{equation}
which is consistent with a power-law in $g(r)$ as given by Eq. (\ref{gr-power}%
). In \cite{Hernan2004} we have considered a delta function followed by a
Taylor expansion around $h=0$ instead of the singular behavior of Eq.
(\ref{ph}). Following the derivations in \cite{Hernan2004} which employ an
effective-medium approximation which assumes an affine motion of the grains
with the external perturbation (being compression of shear) we arrive at the
pressure expressed in terms of the static compressive strain, $\epsilon<0$, as%

\begin{equation}
\sigma= \frac{\phi k_{n}}{6 \pi} [ Z_{c} (-\epsilon)^{3/2} + d_{1}
(-\epsilon)^{5/2-\xi}],
\end{equation}
and the coordination number becomes%

\begin{equation}
Z = Z_{c} +d_{3} \sigma^{(2/3)(1-\xi)}.
\end{equation}
This last result is consistent with Eqs. (\ref{power-1}) and (\ref{power-2})
with $\beta/\alpha= (2/3)(1-\xi)$. Thus a value of $\xi= 0.5$ would fit all
our data. However, we could not find evidence of Eq. (\ref{ph}) in our simulations.%

\begin{figure}
[ptb]
\begin{center}
\includegraphics[
height=2.975in,
width=3.595in
]%
{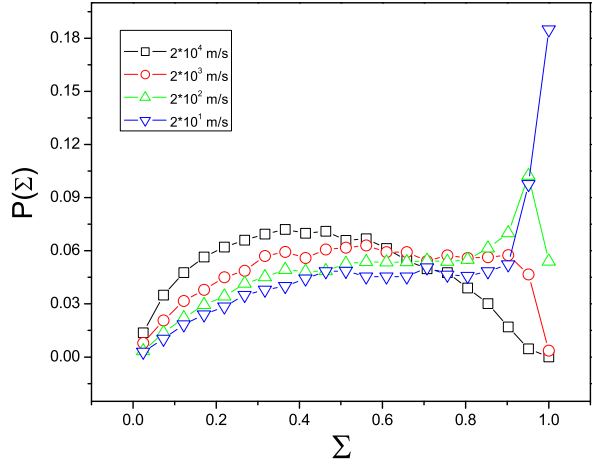}%
\caption{Distributions of $\Sigma$ parameter. The measurements are carried on
4 packings around 650 KPa generated by different compression rates.}%
\label{history-sigma}%
\end{center}
\end{figure}

\begin{figure}
[ptb]
\begin{center}
\includegraphics[
height=2.7856in,
width=3.5198in
]%
{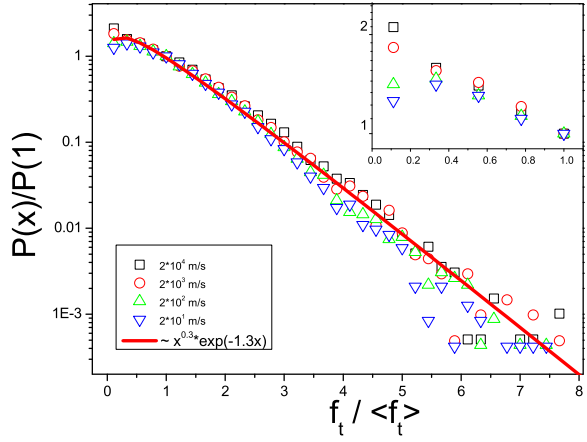}%
\caption{Probability distribution of the tangential forces in packings
generated by different compression rates. The X and Y axes are scaled by the
averaged force and the probability of averaged force respectively. This
scaling convention will be used in all the probability distribution plots of
forces throughout the paper.}%
\label{tangen-pdf}%
\end{center}
\end{figure}

As we discussed in Sec. II, the tangential forces in the Hertz-Mindlin model
are proportional to the normal displacements and therefore depend on the
history of interaction. The history-dependence shown in Fig. \ref{power-laws}c
and \ref{power-laws}d for the granular packings is a result of this
microscopic path-dependence. In our simulations, a lower compression rate
allows the grains to have enough time to relax and slide against each other
during the preparation protocol. This allows for larger tangential
displacement between the grains. On the other hand, a higher compression rate
"freezes" the system so quickly that the grains have no time to slide. Thus we
expect that the tangential force will be smaller in the packing generated by
fast compression rates. In order to quantify this idea, we measured the
plasticity index Eq. (\ref{plasticity}) $\Sigma=\frac{F_{t}}{\mu F_{n}}$ for
four packings around 650 kPa generated by different compression rates. The
probability distributions $P(\Sigma)$, shown in Fig. \ref{history-sigma},
confirm the above argument. In the packing generated by $\gamma=2\times10^{1}%
$m/s, almost $20\%$ of the tangential forces are truncated by the Coulomb
cut-off as evidenced by the fact that $P(\Sigma)$ has a maximum at $\Sigma=1$.
On the other hand, for the fastest compression rate, $\gamma=2\times10^{4}%
$m/s, the peak of the distribution is around $\Sigma\approx0.4$ and
$P(\Sigma=1)\approx0$ implying that there are almost no plastic contacts. This
indicates that the grains in the packing generated by lower compression rates
feel more "sticky" to each other, therefore the packings can sustain non-zero
pressure at lower volume fraction. This explains why $\phi_{c}$ decreases as
$\gamma$ decreases in granular materials. In Fig. \ref{tangen-pdf}, we plot
the probability distribution functions of the tangential forces in the
packings generated by different compression rates. The distributions are
different in the small forces regions. The packing generated by $\gamma
=2\times10^{4}$m/s contains more small forces, which are less than the average
forces, than the packing generated by $\gamma=2\times10^{1}$m/s.%

\begin{figure}
[ptb]
\begin{center}
\includegraphics[
height=2.8193in,
width=3.5328in
]%
{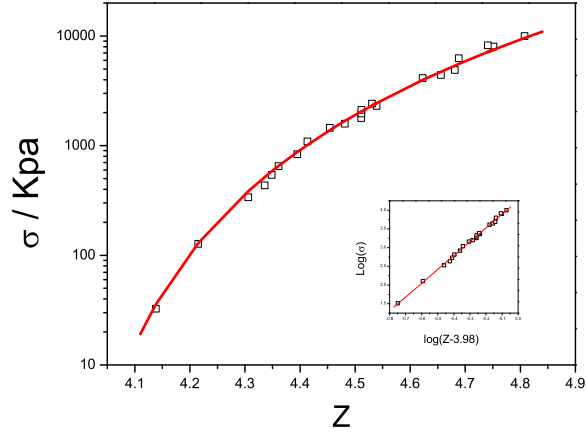}%
\caption{$\sigma$ VS $Z$ in the packings of granular material in the infinite
friction limit. }%
\label{4}%
\end{center}
\end{figure}

We have shown that the critical contact number $Z_{c}\sim4.5$ in the packings
of granular materials with $\mu=0.3$ is larger than the prediction of
isostatic considerations $Z_{c}=4$. We first noticed that $Z_{c}=4$ is
strictly valid when $\mu\rightarrow\infty$. Moreover, we showed above that a
fast compression rate inhibits the developments of tangential forces and
therefore may also contribute to the breakdown of isostaticity. Therefore, we
now investigate the slow compression rate limit of a granular packing in the
$\mu\rightarrow\infty$ limit. We carried out a set of simulations of grains
interacting via Hertz forces Eq. (\ref{gran_normal}) and tangential Mindlin
forces Eq. (\ref{gran-tan}) but without the Coulomb cut-off ($\mu=\infty$) for
a compression rate $\gamma=2\times10^{2}$m/s, slow enough to allow sufficient
time for grains to rearrange. The result of the simulations is shown in Fig.
\ref{4}. The pressure versus volume fraction can be fitted by
\begin{equation}
\sigma\sim\left(  Z-3.98\right)  ^{3.69},
\end{equation}
with a minimal coordination number $Z_{c}=3.98$ indicating the recovery of
isostaticity. We find $\phi_{c}=0.571$ and exponents $\alpha=1.66$ and
$\beta=0.45$. We also find that $Z_{c}$ is larger than $4$ if a compression
rate larger than $\gamma=2\times10^{2}$m/s is used. This may also suggests
that the lack of isostaticity found in \cite{slibertgeo} for frictional
packings might be due to a fast preparation protocol inherent to the
pouring-ball method.%

\bigskip

Finally, we note that, besides $\Sigma$ and the probability distribution
function of tangential forces$,$ we also measure the normal force
distribution, coordination number distribution and radial distribution
function for the four packings around 650 kPa at different compression rates
and found no clear signature of history dependency in these measurements.

\subsection{Micromechanical and microstructural properties}

In order to study the micromechanical and microstructural properties near the
jamming transition we carry out the measurements described in Section II.B.
For the micromechanical properties, we measure the normal and tangential force
distributions \cite{Jasna,liu,slibertforce,antony} and the $\Gamma$
\cite{hernan1} and $\Sigma$ parameters \cite{slibertgeo}. For the
microstructural properties we calculate the distribution of contacts and the
radial distribution function. In this subsection, we will show the results of
the above measurements carried on the packings generated by $\gamma
=1.5\times10^{-3}$m/s for emulsions and $\gamma=2\times10^{4}$m/s for granular
materials. The measurements done on the packings generated by other $\gamma s$
are basically the same.

\subsubsection{Micromechanical properties}%

\begin{figure}
[ptb]
\begin{center}
\includegraphics[
height=3.7593in,
width=4.67in
]%
{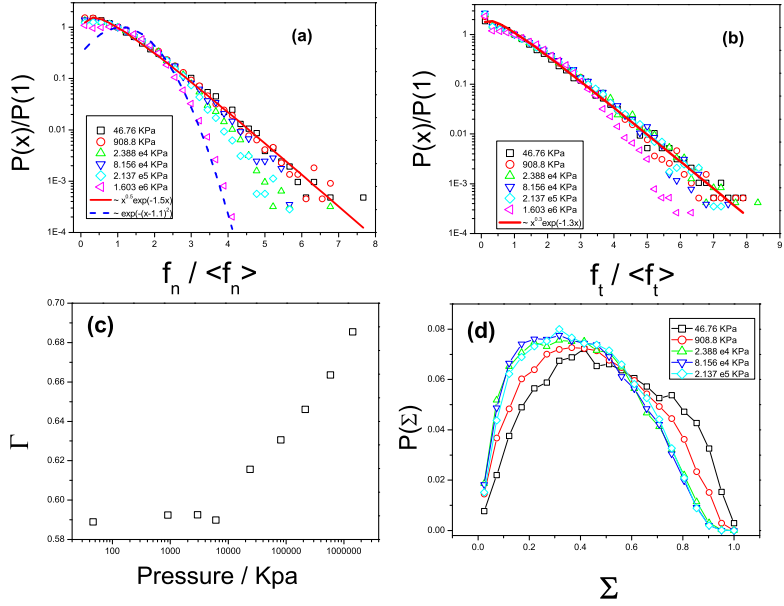}%
\caption{Micromechanical properties at different pressures in the packings of
granular materials. Measurements are carried on the packings generated by
$\gamma=2\times10^{4}m/s$. (a) Distribution of the normal forces, (b)
Distribution of the tangential forces, (c) $\Gamma$ vs pressure and (d)
Distribution of $\Sigma$.}%
\label{press-cont-gran1}%
\end{center}
\end{figure}

\begin{figure}
[ptb]
\begin{center}
\includegraphics[
height=2.6385in,
width=3.2128in
]%
{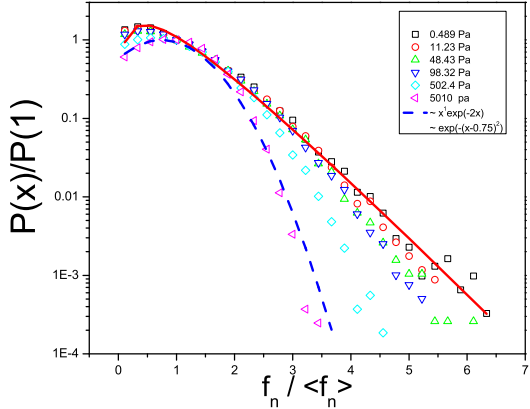}%
\caption{Distribution of the normal forces in emulsions. Measurements were
carried on the packings generated by $\gamma=1.5\times10^{-3}m/s$. }%
\label{press-mech-emul}%
\end{center}
\end{figure}

The normal force distributions in emulsions and granular materials at
different pressures are plotted in Fig. \ref{press-cont-gran1}a and
\ref{press-mech-emul}a respectively as a function of the magnitude of forces
normalized by the average forces. At low pressures, the distributions show a
plateau below the average force which has been considered as the signature of
the jamming transition \cite{ohern1}. The distributions show a broad
exponential tail for forces larger than the average which extend up to eight
times the average force for the packings with the lowest pressures. On the
other hand, $\Gamma$ parameters in Fig. \ref{press-cont-gran1}c and Fig.
\ref{press-mech-emul}b have small values at low pressures, which means the
forces are distributed heterogeneously in space, i.e. they are localized.
Combining the information from the force distributions and the $\Gamma$
parameter we may conclude that at low pressures there are very large forces in
the packings and the forces are distributed heterogeneously. This agrees with
the picture of force chains which have been visualized in granular matter both
in experiment \cite{liu} and simulations \cite{radjai,antony,hernan1}. These
force chains sustain most of the external loading. At low pressures, the
number of the force chains are small and they are well separated
\cite{hernan1}. We also point out that the experiments with confocal
microscopy of compressed emulsions systems \cite{Jasna} did not find evidence
of force chains in the bulk. Our computer simulations of frictionless grains
could not reproduce these results as we find evidence of localization even for
frictionless droplets.

As the pressure increases, the distributions gets narrow; the exponential
tails bend down and transform into a Gaussian-like one, as shown by the
fitting of the dash lines in Figs \ref{press-cont-gran1}a and
\ref{press-mech-emul}a for grains and emulsions. The deviation from the
exponential tail indicates that the very large forces (in comparison with the
average force) are disappearing. At the same time, $\Gamma$ increases sharply
with pressure in Fig. \ref{press-cont-gran1}c and Fig. \ref{press-mech-emul}b,
which indicates that the packing is becoming homogeneous. This sharp increase
happens roughly at 10 MPa for granular materials and 50 Pa for emulsions.
Similar behavior of $\Gamma$ has also been reported in \cite{hernan1} for
grains and the increase of $\Gamma$ was understood as the indication of the
disappearance of well-separated force chains.

The distribution at low pressures, containing a plateau and the exponential
tail, can be fitted by various expressions. Here we choose one from
\cite{Jasna},
\[
P\left(  f\right)  =af^{\lambda}\exp\left(  -(1+\lambda)f\right)  .
\]
This expression comes from a simple Boltzmann equation theory and the power
law coefficient $\lambda$ is determined by the packing geometry of the system.
The fits are shown as solid lines in the plots.

The tangential force distributions in granular materials also show exponential
tails and their pressure dependence is weaker than in the normal force
distributions, as seen in Fig. \ref{press-cont-gran1}b. The distributions of
$\Sigma$ are plotted in Fig. \ref{press-cont-gran1}d for five pressures. As
the pressure decreases, the distribution of $\Sigma$ shifts slightly to higher
values. This is due to the fact that, in our protocol, the system arrests
itself more quickly during the construction of high pressure packings than low
pressure cases. Therefore, the particles in packings at lower pressures have
more time to slide and have larger tangential displacements; therefore larger
tangential forces.%

%

\begin{figure}
[ptb]
\begin{center}
\includegraphics[
height=2.6437in,
width=3.2941in
]%
{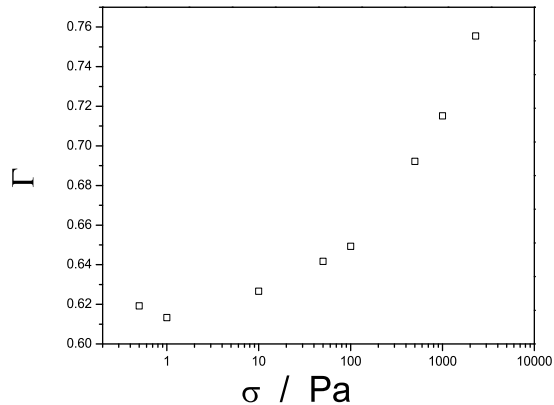}%
\caption{$\Gamma$ vs pressure in emulsions. Measurements were carried on the
packings generated by $\gamma=1.5\times10^{-3}m/s$. }%
\label{press-sig-emul}%
\end{center}
\end{figure}

\subsubsection{Microstructural properties}

\begin{figure}
[ptb]
\begin{center}
\includegraphics[
height=3.8605in,
width=4.7444in
]%
{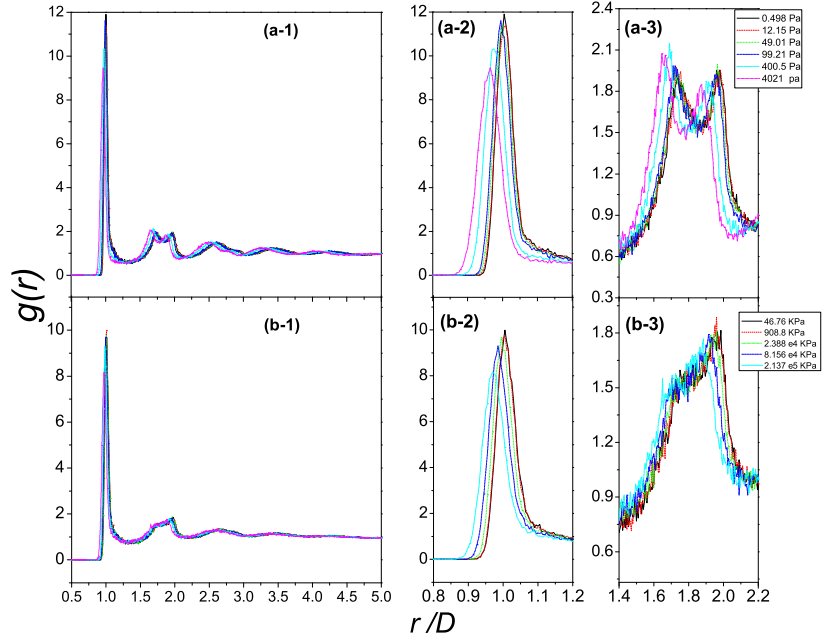}%
\caption{Radial distribution functions (RDF) in the packings of emulsions and
granular materials. (a) the complete RDF in granular matter, (b-1) and (b-2)
are the first and second peak of RDF in granular materials, (c) the complete
RDF in emulsions, (d-1) and (d-2) are the first and second peak of RDF in
emulsions.}%
\label{radial}%
\end{center}
\end{figure}

The radial distribution function (RDF) describes how the particles are
distributed radially. In Fig \ref{radial}, the radial distribution functions
shows a prominent peak at $r=r_{0}$, the second peak at $r=2r_{0}$ and a
sub-peak approximately at $r=\sqrt{3}r_{0}.$ As the system approaches the
jamming transition from high pressures, the distribution functions shift to
the right due to the increase of system size (Figs. \ref{radial}a-1 and
\ref{radial}b-1 for emulsions and grains, respectively). The first peak
decreases in width and increases in hight, as shown in Fig. \ref{radial}a-2
and \ref{radial}b-2. However, we do not find a tendency for the first peak to
evolve into a $\delta-function$ as reported in \cite{ohern2}. In
\cite{slibertgeo} Silbert \textit{et al. } reported the power-law behavior of
Eq. (\ref{gr-power}) near the first peak in granular materials. However, from
our data we can not clearly identify this power-law region. At low pressures
we find $r_{o}\approx2R$, which means that the particles barely touch each other.

\begin{figure}
[ptb]
\begin{center}
\includegraphics[
height=2.7432in,
width=3.6556in
]%
{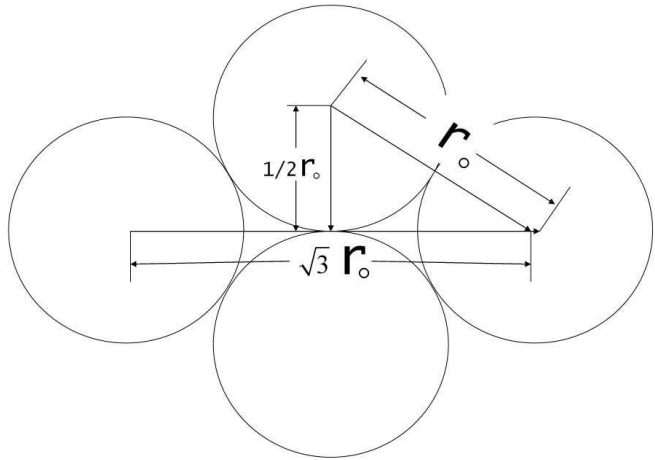}%
\caption{Illustration of possible origin of the sub-peak.}%
\label{hex}%
\end{center}
\end{figure}

The splitting of the second peak in the RDF observed in Fig. \ref{radial}b-2
and Fig. \ref{radial}b-3 has been reported by various authors studying dense
liquids and granular materials \cite{truskett2,clarke,Kob,H.Li}. The authors
of Ref. \cite{truskett2} consider this splitting to be a structural precursor
to the freezing transition and the development of long-range order. They also
attribute this splitting to four-particle hexagonally close-packed
arrangements. As shown in Fig. \ref{hex}, the second nearest separation in the
hexagonal packings is $\sqrt{3}r_{o}$. Therefore the splitting sub-peak
possibly indicates the appearance of small hexagonal clusters like the one in
the figure.%

%

\begin{figure}
[ptb]
\begin{center}
\includegraphics[
height=2.7034in,
width=3.2932in
]%
{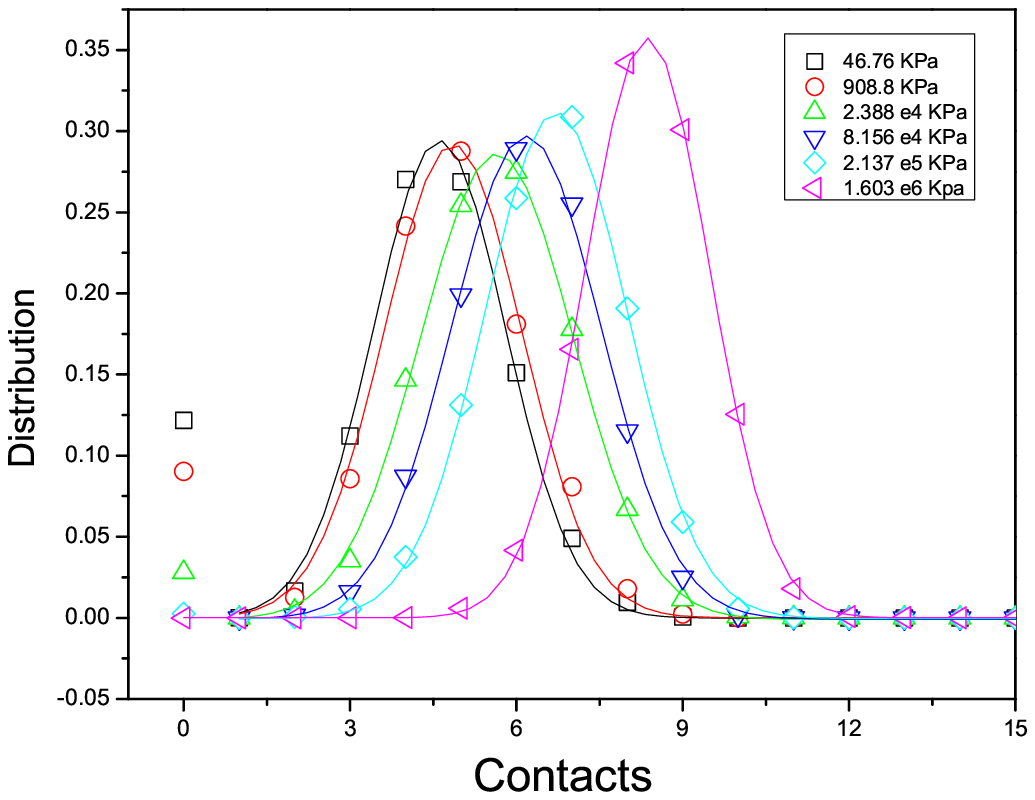}%
\caption{Distribution of contacts in the packings of granular materials. The
lines are Guassian fits to the distributions, excluding the floaters.}%
\label{press-cont-gran}%
\end{center}
\end{figure}

The distributions of contacts are shown in Fig. \ref{press-cont-gran}a for
granular materials and in Fig. \ref{press-cont-gran}b for emulsions. Both
distributions can be fitted by Gaussians. As the pressure is increased the
distributions move to the right and get slightly narrower. The floaters (those
particles with zero coordination number) exist in granular materials at low
pressure (around $12\%$ at $46.74$ $kPa$) while the fraction of floaters is
much lower ($2\%$ at $0.498$ $Pa$) in emulsions. As expected the number of
floaters decreases as pressure increases. Excluding the floaters, we can fit
the distributions with a Gaussian function, as shown by lines in Fig.
\ref{press-cont-gran}a and \ref{press-cont-gran}b.
%

\begin{figure}
[ptb]
\begin{center}
\includegraphics[
height=2.8928in,
width=3.3788in
]%
{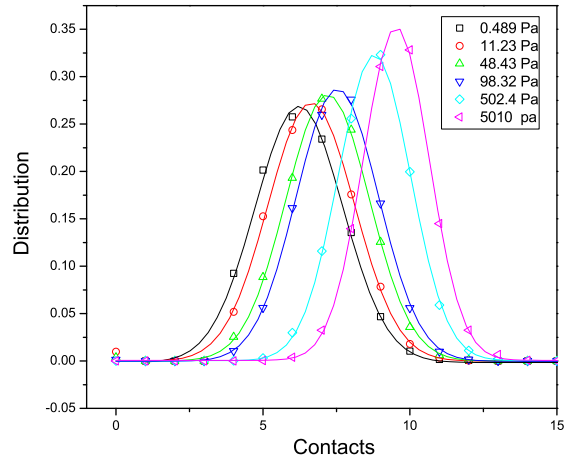}%
\caption{Distribution of contacts in the packings of emulsions. The lines are
Guassian fits to the distributions, excluding the floaters.}%
\label{press-cont-emul}%
\end{center}
\end{figure}

\section{Summary}

We have studied the jamming transition in packings of emulsions and granular
materials via molecular dynamics simulations. Power-law scaling is found for
the vanishing of the pressure and excess number of contacts as the system
approaches the jamming transition from high volume fractions. The emulsion
system jams at
$\phi_{c}=0.645$ independent on the construction histories while granular
materials jam at $\phi_{c}$ between $0.576$ and $0.604$ depending on the
construction histories. We found that the preparation protocol has strong
effect on the tangential forces in granular materials. Longer construction
times of the packings allows the particles to relax and slide against each
other and therefore to have larger tangential displacements, which lead to
larger tangential forces. Isostaticity is found in the packings close to the
jamming transition in emulsions and in granular materials in the limit of
infinite friction and slow compression rate. Heterogeneity of the force
distribution increases while the system approaches the jamming transition,
demonstrated by the exponential tail in the force distributions and the small
values of the participation number. However, no signatures of jamming
transitions are observed in structural properties, like the radial
distribution functions and the distributions of contacts.

Acknowledgments: We acknowledge financial support from the Department of
Energy, Division of Basic Science, Division of Materials Sciences and
Engineering, DE-FE02-03ER46089, and the National Science Foundation, DMR
Materials Science Program, DMR-0239504.

\end{document}